\documentclass[aps,prd,12pt]{revtex4}

\usepackage{amsmath,amssymb}

\usepackage{color}
\usepackage[dvips]{graphicx}

\newcommand{\lsim}{\mathrel{\mathop{\kern 0pt \rlap
  {\raise.2ex\hbox{$<$}}}
  \lower.9ex\hbox{\kern-.190em $\sim$}}}
\newcommand{\gsim}{\mathrel{\mathop{\kern 0pt \rlap
  {\raise.2ex\hbox{$>$}}}
  \lower.9ex\hbox{\kern-.190em $\sim$}}}

\newcommand{\be}{\begin{equation}}
\newcommand{\ee}{\end{equation}}
\newcommand{\beqarr}{\begin{eqnarray}}
\newcommand{\eeqarr}{\end{eqnarray}}

\begin{document}

\preprint{KIAS-P10026}

\title{Leptogenesis origin of Dirac gaugino dark matter}



\author{Eung Jin Chun}
\affiliation{Korea Institute for Advanced Study\\
Hoegiro 87, Dongdaemun-gu, Seoul 130-722, Korea\\
Email) {ejchun@kias.re.kr}
 }

%

\begin{abstract}
The Dirac nature of the gauginos (and also the Higgsinos) can be
realized in  $R$-symmetric supersymmetry models. In this class of
models, the Dirac bino (or wino) with a small mixture of the Dirac
Higgsinos is a good dark matter candidate. When the seesaw
mechanism with Higgs triplet superfields is implemented to account
for the neutrino masses and mixing, the leptogenesis driven by the
heavy triplet decay is shown to produce not only the
matter-antimatter asymmetry but also the asymmetric relic density
of the Dirac gaugino dark matter. The dark matter mass turns out
to be controlled by the Yukawa couplings of the heavy Higgs
triplets, and it can be naturally at the weak scale for a mild
hierarchy of the Yukawa couplings.
\end{abstract}


\maketitle


The gauge sector of the supersymmetric standard model may have an
extended structure beyond the conventional  N=1 supersymmetry to
realize the (almost pure) Dirac nature as in the N=2
supersymmetry \cite{N=2}.  To maintain the Dirac structure, one
needs an $U(1)_R$ symmetry which forbids the conventional $\mu$
and $A$ terms \cite{hall91}. This also requires an extended Higgs
sector \cite{kribs07,chun09}. In a more minimal Dirac gaugino
scheme with the conventional Higgs sector \cite{benakli08}, the
$R$ symmetry is broken by the presence of the $\mu$ and $B\mu$ (or
$A$) terms. Thus, although the Dirac structure can be impose at
tree level, one loop corrections involving these $R$ breaking
terms induce sizable Majorana mass terms.

The Dirac gaugino/Higgsino models  have some interesting
phenomenological consequences. First of all, the Dirac or Majorana
nature of the gauginos can be tested by several observables at the
LHC \cite{nojiri07,plehn08,choi08}. Secondly, the Dirac gaugino denoted by
$\chi$ (assumed to be mostly Dirac bino in this work) is a good
dark matter (DM) candidate \cite{ullio06,hsieh07,benakli09,chun09} as
in the usual minimal supersymmetric standard model, while the
Dirac Higgsino dark matter is ruled out by the direct detection
experiment as well as the cosmic antiproton measurement
\cite{chun09}. This is due to the fact that the Dirac Higgsino has
a vector interaction with the $Z$ boson leading to a large
spin-independent cross-section with nuclei, or a large
annihilation to $W^+ W^-$. Thereby, one can derive a limit on the
Higgsino mixture (denoted by $\delta_V$) allowing the vector
interaction, $\delta^2_V \bar{\chi}\gamma_\mu \chi Z^\mu$: $\delta_V <
0.2-0.3$ depending on the dark matter mass.  Another distinctive
feature of the Dirac gaugino is that its annihilation into
fermion--anti-fermion through the t-channel sfermion exchange,
$\chi\bar{\chi} \to f\bar{f}$, possesses neither the chirality
(fermion mass) suppression nor the velocity suppression.
Consequently, the Dirac gaugino annihilation
can be too strong to produce a sizable thermal relic density in
the standard cosmology if the corresponding sfermion mass is
light, or the Higgsino component is on the border of the current limit.
In this case, a non-standard cosmology or a non-thermal production has to be
assumed to yield the right relic density of the Dirac gaugino dark
matter.

\bigskip

Concerning the origin of the dark matter density, there appears an
intriguing relation of the dark matter and baryon densities:
$\Omega_{DM}/\Omega_B \simeq 5$ \cite{wmap5}. The coincidental
feature of $\Omega_{DM} \sim \Omega_B$  would imply that they
originated from the same source, and it led to a lot of ideas on
how such a property can be realized in Nature \cite{nussinov85}.
In this paper, we will show that the conventional leptogenesis
\cite{fukugita86} in the type II seesaw, which introduces
$SU(2)_L$ triplets to account for the
neutrino masses and mixing \cite{davidson08}, can be the common
origin of the Dirac gaugino dark matter and baryon densities.
In other words, when
the leptogenesis produces a $B-L$ asymmetry, $Y_{B-L}\equiv
(n_B-n_{\bar{B}})/s$ where $s$ is the entropy density, a
non-vanishing asymmetry in the Dirac dark matter,
$Y_{DM}\equiv (n_{\chi} -n_{\bar{\chi}})/s$, is necessarily produced. As
will be shown explicitly, this can be understood by the
presence of the ``dark matter number" or the ``Dirac charge"
\cite{choi08} which is associated with the conventional $R$
symmetry.  The Dirac charge is conserved by all the interactions of  the
supersymmetric standard model excluding the type II seesaw sector
\cite{note1}.  As a consequence, the dark matter and baryon
densities follow the well-known relation:
\begin{equation} \label{ratio-DM-B}
 {\Omega_{DM}\over \Omega_B} \approx {m_{DM} Y_{DM}
 \over m_n Y_{B-L} } \approx 5 \,.
\end{equation}
where $m_n$ is the nucleon mass. If $Y_{DM}\sim Y_{B-L}$,
Eq.~(\ref{ratio-DM-B}) requires  a light dark matter mass
$m_{DM}\sim 5$ GeV.  However, we find that the ratio,
$Y_{DM}/Y_{B-L}$, is controlled by the branching ratios of the
heavy scalar or fermion triplets, and can be in the range of
$0.1-0.01$. Thus, the dark matter mass can be naturally at the
weak scale in our mechanism.

\bigskip

Let us begin our main discussion by writing down the
superpotential of an $U(1)_R$ model with an extended  gauge-Higgs
sector \cite{kribs07, chun09} as well as the type II seesaw
sector.  For this, we introduce (anti) Higgs doublet superfields
$R_{1,2}$,  anti-gaugino multiplet superfields $\Phi^a$, and Dirac
pairs of the $SU(2)_L$ triplet superfields $(\Delta_i,
\Delta^c_i)$ in addition to the usual minimal supersymmetric
standard model superfields.  Furthermore, supersymmetry and $R$
symmetry are supposed to be broken by  a constant superpotential
$W_0$ and a superfield $X$ which takes the vacuum expectation
value, $\langle X \rangle = \theta^2 F_X$. We take the typical
size of  the supersymmetry breaking terms $F_X \sim W_0 \sim
\tilde{m}$ with $\tilde{m}\sim1$ TeV in the unit of the Planck
mass: $M_P=1$.

 Assigning the
following $R$ charges
\begin{equation} \label{Rcharge}
 \begin{array}{|cccccc|cc|}
 \hline
 L & H_{1,2} & R_{1,2} & \Phi^a & \Delta & \Delta^c &  X & W_0 \cr
 \hline
 {1\over2} & {1\over2} & -{3\over2} & 3 & 1 & 1 & -1 & 2 \cr
 \hline
 \end{array}
\end{equation}
one can write down the new gauge and $R$ invariant superpotential
terms given by
\begin{eqnarray}
 W_{R}&=&\mu_1 H_1 R_2 + \mu_2 H_2 R_1 -\sqrt{2} g_a (\xi_1 H_1 T^a R_2
 +\xi_2 H_2 T^a R_1) \Phi^a + h_a X W_a^\alpha Q_\alpha \Phi^a\,,
  \nonumber \\
 W_{II} &=& y_i \Delta_i L L + \lambda_{i} \Delta_i H_1 H_1
  + \lambda^c_{i} \Delta^c_i H_2 H_2 + M_i \Delta_i \Delta^c_i \,.
\label{WRII}
\end{eqnarray}
Here $W_a^\alpha$ are the usual gauge superfields, $Q_\alpha$ is
the covariant supersymmetric derivative, and $L$ and $H_{1,2}$ are
the lepton and Higgs doublet superfields, respectively (we
suppress the family index of $L$). [Remark that more stringent $R$
symmetries defined in Ref.~\cite{chun09} do not allow the
superpotential $W_{II}$ in Eq.~(\ref{WRII}).] From the last term
of the superpotential $W_R$ in Eq.~(3), one gets the Dirac gaugino
mass term $M_D^a \lambda_a \lambda^{c\,a}+h.c.$ where $M_D^a = h_a
\langle F_X \rangle$.  Here let us note that we can define the
notion of the DM number or the Dirac charge \cite{choi08} as
follows:
\begin{equation}
 \begin{array}{|cc|cccccc|cccc|}
 \hline
 \lambda^a & \lambda^{c\,a} & \, f & \tilde{f} & \, f^c & \tilde{f}^c
 & h_i & \tilde{h}_i &
  \, \delta & \tilde{\delta} & \delta^c & \tilde{\delta}^c \cr
 \hline
 1 & -1 & 0 & 1 & 0 & 1 & 0 & -1 & 0 & -1 & 2 & 1 \cr
 \hline
 \end{array}
\end{equation}
where $\tilde{f}$ denotes the scalar partner of the quark or
lepton field $f$, $\tilde{h}_i$ the fermion partner of the Higgs
boson $h_i$, and $\tilde{\delta}$/$\tilde{\delta}^c$ is the
fermionic counterpart of the triplet/anti-triplet boson
$\delta$/$\delta^c$.  The Dirac charge conservation is respected
by all the usual gauge and Yukawa interactions except the
$\lambda_i$ coupling.  That is, the DM number is broken in the
decays of the triplets and this is why a non-vanishing DM
asymmetry $Y_{DM}$ can be generated along with the $B-L$ asymmetry
$Y_{B-L}$ as will be explicitly shown later.

The supersymmetry breaking field $X$ gives the usual soft masses
to sfermions and Higgs bosons through
 $$X X^\dagger L L^\dagger|_D, \quad\mbox{etc.}.$$
Consistently with the $R$ symmetry in Eq.~(\ref{Rcharge}), one
finds that  the extended $\mu$ parameters and the Higgs boson $B$
term come from the following $R$ invariant terms
\begin{equation} \label{R-mass}
 H R X^\dagger\big|_D\quad \mbox{and}\quad
 H_1 H_2 X W\big|_F \,,
\end{equation}
respectively.  In addition to these, the following operator are
also allowed;
\begin{equation} \label{nonR-mass}
 (H_1 H_2)^2\big|_F,\quad
 R_1 R_2 W X^\dagger\big|_D
\end{equation}
which spoil the Dirac structure by inducing the usual Higgsino mass terms:
$\delta\mu_H {H}_1{H}_2$ and $\delta\mu_R {R}_1 {R}_2$ where $\delta \mu_H \sim
\langle H_1 H_2 \rangle/M_P$ and $\delta \mu_R \sim \tilde{m}^2/M_P$.
From Eqs.~(\ref{WRII}, \ref{R-mass}, \ref{nonR-mass}), we get   the
almost $R$ symmetric gaugino-Higgsino mass matrix:
\begin{equation}
\label{mass-matrix}
{\mathcal{M}} \!\! =\!\!
\left[\begin{array}{c c c c c c}
 0 & M_D & 0 &  -\xi_1 m_Z s_W c_\beta & 0 & \xi_2 m_Z s_W s_\beta  \\
 M_D & 0   & -m_Z s_W c_\beta & 0 &  m_Z s_W s_\beta & 0  \\
 0  & -m_Z s_W c_\beta & 0 & \mu_1 & \delta\mu_H & 0  \\
 -\xi_1 m_Z s_W c_\beta & 0 & \mu_1 & 0 & 0 & \delta\mu_R \\
  0  &  m_Z s_W s_\beta & \delta\mu_H & 0 & 0 & \mu_2   \\
 \xi_2 m_Z s_W s_\beta & 0 & 0 & \delta\mu_R & \mu_2 & 0\\
\end{array}\right]
\end{equation}
in the Weyl basis of $(\tilde{B}^c, \tilde{B}, \tilde{h}_1,
\tilde{r}_2, \tilde{h}_2, \tilde{r}_1)$. Here we included only the
bino $\tilde{B}$  and anti-bino $\tilde{B}^c$, decoupling the wino
components for simplicity. One can see that the above mass matrix
maintains the Dirac structure except the $\delta \mu_{H,R}$ terms.
When $M_D \ll \mu_{1,2}$, the lightest component is mostly the
Dirac bino, $\chi\approx (\tilde{B}, \bar{\tilde{B}}^c)$, with
$m_\chi\approx M_D$, and it can have a small mixture of the two
Dirac Higgsinos, with the mixing parameter $\delta_V \sim m_Z
s_W/\mu_{1,2}$, which was used previously.

The mass terms $\delta \mu_{H,R}$ yield small  Majorana masses $\delta m_M$
of the  bino and anti-bino components given by
\begin{equation} \label{mM}
 \delta m_M \sim \delta \mu_{H,R}{ m_Z^2 s_W^2 \over \mu_1\mu_2} .
\end{equation}
This splits the Dirac bino into two Majorana mass eigenstates with
a small mass difference of $\delta m_M$.  In the presence of such
a  mass splitting, there occurs an oscillation of the bino to the
anti-bino state with the oscillation time $t_{osc}\sim 1/\delta
m_M$, which erases out the bino--anti-bino asymmetry produced via
leptogenesis. Then, the usual interaction like $\chi\bar{\chi}\to
f\bar{f}$ determines the relic abundance by the standard
freeze-out process, independently of the primordial asymmetric
relic density. This situation can be avoided if the
bino--anti-bino oscillation starts after the freeze-out of
annihilation processes: $t_{osc}> t_{f.o.}$.  It implies $T_{osc}
< T_{f.o.}$ where the two temperatures are given approximately by
$T_{osc}\sim \sqrt{\delta m_M M_P}$ and $T_{f.o.}\sim m_{DM}/x_f$
with $x_f \sim 20$.  Thus, we get the condition:
\begin{equation} \label{mM-osc}
 \delta m_M \lesssim {1 \over x_f^2} {m_{DM}^2 \over M_P}
\end{equation}
which is of the order (\ref{mM}) and can be readily satisfied
with $m_Z \ll \mu_{1,2}$ and a bit suppressed $\delta \mu_{H,R}$.
Let us remark that the heavier Majorana state $\chi_2$ decays to
the lighter one $\chi_1$ through a radiative decay $\chi_2 \to
\chi_1 \gamma$ induced by one-loop corrections, and its lifetime
is shorter than the age of the universe given the mass splitting
of the order (\ref{mM}) \cite{chun09}. Therefore, the dark matter
today consists of the Majorana fermion $\chi_1$.

\bigskip

In order for the leptogenesis to be the main source of the
(asymmetric) Dirac gaugino dark matter population, the standard
thermal freeze-out process should be strong enough to suppress the
usual (symmetric) dark matter population.  As alluded in the
introduction, the gauge annihilation into a fermion-antifermion
pair, $\chi \bar{\chi} \to f \bar{f}$ (mediated by the $t$-channel
sfermion), does not contain the chirality suppression contrary to
the Majorana dark matter case and thus this can be sufficiently
strong if the corresponding sfermion is not too heavy. Taking the
final state fermion as a lepton, we get the annihilation rate
\cite{chun09}:
\begin{equation}
 \langle \sigma v \rangle = \sum_l {2\pi\alpha^2 \over c_W^4}
 {m_\chi^2 \over (m^2_{\tilde{l}} + m^2_\chi)^2}
\end{equation}
where $\alpha$ is the fine-structure constant, $c_W$ is the cosine
of the weak mixing angle, and $m_{\tilde{l}}$ is the slepton mass of each lepton flavor $l$.  
Taking almost degenerate three slepton masses and 
$m_{\tilde{l}} \approx m_\chi$, one finds
\begin{equation}
 \langle \sigma v \rangle \approx 1.8\times10^{-9}\,
 \mbox{GeV}^{-2} \left( 420 \mbox{ GeV} \over m_{\chi}\right)^2
 \,.
\end{equation}
As the dark matter relic density follows the relation:
$\Omega_\chi h^2 \approx 2\times10^{-10} \mbox{GeV}^{-2} /\langle
\sigma v \rangle$, we conclude that the symmetric relic component
is suppressed below the observed value if $m_\chi \lesssim 420$
GeV.  Such an upper bound on the lightest supersymmetric
particle could be tested in the near future as the
current LHC data constrains the significant parameter space of
the constrained minimal supersymmetric standard model based on 
the minimal supergravity breaking scenario \cite{cms11}. Note, however, that the experimental limit strongly depends on the supersymmetry breaking scenario and can be easily loosened in models with non-universal gaugino masses for instance.

\bigskip

The type II seesaw mechanism gives rise to the neutrino mass
through the vacuum expectation values of the triplets:
\begin{equation}
 m_\nu = \sum_i y_i \langle \Delta^0_i \rangle = \sum_i y_i \lambda^c_i
 {\langle H_2^0 \rangle^2 \over M_i }
\end{equation}
for $\langle H_2^0 \rangle \ll M_i$.  Thus the product of the two
Yukawa couplings is required to be
\begin{equation} \label{ylambda}
 \sum_i y_i \lambda^c_i \left({10^{12} \mbox{ GeV} \over M_i}\right)
 \approx 2\times10^{-3}
\end{equation}
for the atmospheric neutrino mass scale $m_\nu = \sqrt{\Delta
m^2_{atm}} \approx 0.05$ eV and $\langle H^0_2 \rangle =174$ GeV.

Given the neutrino mass condition (\ref{ylambda}), let us now
discuss how successfully leptogenesis can arise. The type II
leptogenesis  requires  more than two heavy triplets and the CP
violating decays occur through one-loop self-energy diagrams
\cite{ma98}.  The type II leptogenesis differs from the type I
leptogenesis as the former uses the decay of the gauge charged
particle, $\Delta$, which interacts not only through the Yukawa
coupling but also through the gauge coupling. However, the resulting asymmetry depends mildly on the gauge interaction 
and the gauge interaction effect becomes irrelevant
for high enough triplet masses in a strong washout regime \cite{chun0510}.
Furthermore, a successful leptogenesis
is shown to arise for $M_i > 10^{11}$ GeV \cite{hambye05}.  
In the below, we will illustrate that there exits an appropriate parameter region where both leptogenesis and dark matter genesis work successfully. 

From the decay of the lighter triplet $(\Delta_1, \Delta^c_1)$ with the mass $M_1$, the lepton
(or $B$-$L$) and CP asymmetry arises as
\begin{equation}
 \varepsilon_{B-L} \approx {3\over 16\pi } { \mbox{Im}[ (y_1
 \lambda_1^c) (y_2 \lambda_2^c)^*] \over |y_1|^2 + |\lambda^c_1|^2 }
 {M_1 \over M_2}
\end{equation}
in the limit of $M_1 \ll M_2$ \cite{hambye05}.  Here we neglect
the contribution of $\lambda_i$ assuming they are smaller than the
other Yukawa couplings as will be required in the following
discussions, and we will also  assume  $|y_i| \sim |\lambda_i^c|$.
Then,  one gets
\begin{equation} \label{CPA}
 \varepsilon_{B-L} \sim {3\over 16\pi} |y_2||\lambda^c_2| \sin\phi {M_1
 \over M_2} \sim 10^{-5} \sin\phi
 \left( M_1 \over 10^{12} \mbox{ GeV}\right)
 \left( 10 M_1 \over M_2 \right)
\end{equation}
where we have used Eq.~(\ref{ylambda}) and $\sin\phi$ is the
amount of CP phase contribution.  For considering the
out-of-equilibrium condition, it is convenient to calculate the
$K$-factor:
\begin{equation}
 K \equiv {\Gamma_1 \over H} \approx 32  \left( \tilde{m}_1 \over 0.05
 \mbox{ eV} \right)
\end{equation}
where $\Gamma_1 \approx (|y_1|^2+|\lambda^c_1|^2) M_1/8\pi$, and $H$ is
the Hubble parameter at the temperature $T=M_1$. Here the effective neutrino mass $\tilde{m}_1 \equiv |y_1 \lambda_1| \langle H_2^0\rangle^2/M_1$ does not have to be related with the true neutrino mass in Eq.~(12) as it is the sum of the two contributions.  Thus, $K$ can take any value generically smaller than 32.
Nevertheless, we will take $K=32$ to show that a successful leptogenesis can arise 
even in this strong wash-out regime driven by a large
inverse-decay rate. The efficiency factor for large $K$ can be
approximated as $\eta \approx 1/1.2 K (\ln K)^{0.8}$
\cite{buchmuller04}, and thus one finds a small wash-out factor of 
\begin{equation} \label{etamax}
  \eta  \approx \left.{1\over 1.2 K (\ln K)^{0.8} }
 \right|_{K=32}
 \approx 10^{-2}.
\end{equation}
Of course, the wash-out effect can become smaller for smaller $K$ taking smaller effective neutrino mass $\tilde{m}_1$.
Putting together the values of the CP asymmetry (\ref{CPA}) and
the efficiency factor (\ref{etamax}), we finally obtains the
resulting baryon asymmetry:
\begin{equation}
  Y_{B-L} \sim 10^{-2} \varepsilon_{B-L} \eta \lesssim 10^{-9}
  \left( M_1 \over 10^{12} \mbox{ GeV}\right)
  \left( 10 M_1 \over M_2 \right)
\end{equation}
which shows that the observed value of $Y_{B-L}\approx 10^{-10}$
can be obtained for $M_i\gtrsim 10^{11}$ GeV with $\sin\phi\sim
1$.

\bigskip

Let us finally show that the type II leptogenesis can naturally
generate the right density of the asymmetric dark matter whose
mass is at the electroweak scale. For the simplicity of our
discussion, we will suppress the family indices of the triplets
unless otherwise necessary. 
 To verify the generation of 
  the Dirac dark matter asymmetry together with the $B-L$ asymmetry
coming from the CP-violating
decays of the scalar and fermion triplets, we list the relevant
two-body decays of the scalar triplets ($\delta$, $\delta^c$), and 
the Dirac fermion triplet $(\tilde{\delta}, \bar{\tilde{\delta}}^c)$:
\begin{equation} \label{II-decay}
 \begin{array}{rllll}
 \bar{\delta} &\to&
 l l, &
 \bar{h}_2 \bar{h}_2, &
 \tilde{h}_1 \tilde{h}_1 \; ({\chi}^c {\chi}^c) \cr
 \delta^c &\to&
 \tilde{l} \tilde{l}\; (\chi\chi),&
 \bar{\tilde{h}}_2 \bar{\tilde{h}}_2\; (\chi\chi),&
 h_1 h_1 \cr
 (\bar{\tilde{\delta}}, \tilde{\delta}^c) & \to&
 l \tilde{l}\; (\chi), &
 \bar{h}_2 \bar{\tilde{h}}_2\; (\chi),&
 h_1 \tilde{h}_1 \; ({\chi}^c)
 \end{array}
\end{equation}
where we showed the resulting gaugino ($\chi$) or anti-gaugino ($\chi^c$)
dark matter production in the parentheses 
which comes from the decays of Higgsinos or slpetons:
$\tilde{h}_i \to h_i \chi^c$ and $\tilde{l}\to l \chi$, {\it etc}.

For our discussion, we need to remind some general property of
CP asymmetries in an $X$ particle decay. The CPT
conservation and unitarity tell us that
\begin{equation} \label{cpt-unitary}
 \sum_{j \in all}\Gamma(X\to j  ) = \sum_{j\in all} \Gamma(\bar{X} \to j)
\end{equation}
where the summation is over all the possible final states $j$. For
the CP asymmetry for a particular decay channel, $X \to j$, we
define
\begin{equation}
 \varepsilon_X^j \equiv {\Gamma(X\to j)- \Gamma(\bar{X}\to \bar{j})
 \over \Gamma_X^{tot} }
\end{equation}
with $\Gamma_X^{tot}= \sum_{k \in all} \Gamma(X\to k)$.
Eq.~(\ref{cpt-unitary}) implies the obvious identity:
\begin{equation} \label{all-epsilon}
\sum_{j \in all} \varepsilon^j_{X} =0 \,.
\end{equation}
However, a summation over a subset of all the final states needs
not to vanish.  In leptogenesis, one gets a non-zero $B-L$
asymmetry, $\varepsilon_{B-L} \equiv  \sum_{j\in B-L} Q^{B-L}_j\,
\varepsilon^j_X$, for the final states carrying the $B-L$ number
$Q^{B-L}_j$.   Likewise, we can also obtain a non-vanishing
asymmetry of the Dirac gaugino dark matter, $\varepsilon_{DM}
\equiv \sum_{j\in DM} Q^{DM}_j \,\varepsilon^j_X$, for the final
states $j$ with the DM number $Q^{DM}_j$.
We can now apply the above properties in the processes of
Eq.~(\ref{II-decay}).  The $B-L$ and DM asymmetries are given by
\begin{eqnarray}
 \varepsilon_{B-L} &=& -2\, [ \varepsilon^{ll}_{\bar{\delta}}
                     +  \varepsilon^{\tilde{l}\tilde{l}}_{\delta^c}
                     + \varepsilon^{l\tilde{l}}_{\tilde{\delta}^c}]
 \nonumber\\
 \varepsilon_{DM} &=& -2\, [ \varepsilon^{\tilde{h}_1 \tilde{h}_1}_{\bar{\delta}}
                     +  \varepsilon^{h_1 h_1}_{\delta^c}
                     + \varepsilon^{h_1\tilde{h}_1}_{\tilde{\delta}^c}]
\end{eqnarray}
where the expression for $\varepsilon_{DM}$ is obtained
using the identity (\ref{all-epsilon}) in the leading order.
Including the contribution of $\lambda_i$ neglected in Eq.~(14),
 we obtain the
following CP asymmetries in the $(\Delta_1 , \Delta^c_1)$ decays:
\begin{eqnarray}
 \varepsilon_{B-L} &=& -6\, [ \epsilon_1 +\epsilon_2 ]
 \nonumber\\
 \varepsilon_{DM} &=& +6 \, [ \epsilon_1 +\epsilon_{3} ] \,.
 \label{epsilonBLDM}
\end{eqnarray}
Here the quantities $\epsilon_{1,2,3}$ are proportional to
different CP phases given by
\begin{eqnarray}
 \epsilon_1 &\propto& \mbox{Im}[y_1 y_2^* \lambda_{1}^*
 \lambda_{2}] \nonumber\\
 \epsilon_2 &\propto& \mbox{Im}[y_1 y_2^* \lambda^c_{1}
 \lambda_{2}^{c*} ]
 \label{epsilon123} \\
 \epsilon_{3} &\propto& \mbox{Im}[\lambda_{1}^{*}
 \lambda_{2} \lambda_{1}^{c*} \lambda_{2}^c] \,. \nonumber
\end{eqnarray}
Note that the dark matter asymmetry vanishes if we put
$\lambda_{i}=0$, which was obvious from the identity
(\ref{all-epsilon}) applied to the precesses (\ref{II-decay}).
This can also be understood by considering the notion of the DM
number or the Dirac charge \cite{choi08} defined in Eq.~(4).
  It shows that the DM number is broken by the coupling $\lambda_i$
in the decays of the triplets and this is why a non-vanishing DM
asymmetry $Y_{DM}$ can be generated along with the $B-L$ asymmetry
$Y_{B-L}$. Applying the above arguments, one can confirm that  no
dark matter  asymmetry is generated in the type I (and also III)
seesaw mechanism introducing singlet right-handed neutrinos
(fermion triplets) as mentioned in the introduction.

Eq.~(\ref{epsilonBLDM}) gives
\begin{equation}
 {\Omega_{DM} \over \Omega_B} \approx { m_{\chi} |\epsilon_1 +
 \epsilon_3| \over m_n |\epsilon_1 +\epsilon_2| } \approx 5
\end{equation}
and again the ratio $|Y_{DM}/Y_B|=|\epsilon_1 +
 \epsilon_3|/|\epsilon_1 +\epsilon_2|$ becomes vanishingly small
in the limit of $\lambda_i \to 0$. Therefore, the dark matter mass
is determined by the hierarchy of the Yukawa couplings $y,
\lambda$ and $\lambda^c$ (the family indices suppressed). Writing
down schematically $|\epsilon_1 + \epsilon_3| \sim \lambda^2
(y^2+\lambda^{c2})$ and $|\epsilon_1 + \epsilon_2| \sim y^2
(\lambda^2+\lambda^{c2})$, one finds that $m_{DM} \sim 5
(y^2/\lambda^2)$ GeV for $\lambda < y \sim \lambda^c$. A mild
hierarchy of $\lambda/y \sim 0.3-0.13$ sets the dark matter mass
at the weak scale: $m_{DM} \sim 50-300$ GeV.

\bigskip

A comment is in order concerning the dynamics of generating the
lepton and dark matter asymmetries. As discussed already, the
heavy decaying particles, $(\Delta, \Delta^c)$, are charged under
the Standard Model gauge symmetries and thus they can  annihilate
through the (strong) gauge interactions which becomes important
for the low seesaw mass scale $M_i$. In this case,
 $Y_{B-L}$ and $Y_{DM}$ can be much more suppressed by a large Boltzmann factor
and thus the CP asymmetries, $\varepsilon_{B-L}$ and
$\varepsilon_{DM}$, need to be much larger than calculated before.
This may be achieved by the resonant enhancement
\cite{resonant96}. Then, the triplet mass could be as low as
${\cal O}(\mbox{TeV})$ allowing maximum CP asymmetries:
$\varepsilon_{B-L, DM} \sim {\cal O}(1)$. It is interesting to
probe such a TeV scale leptogenesis with a large CP asymmetry at
colliders by observing the lepton asymmetry, $N(l^+ l^+) - N(l^-
l^-)$, in the triplet decay of $\Delta^{\pm\pm} \to l^\pm l^\pm$
\cite{chun0508}.

\bigskip

To conclude, we have shown that the usual leptogenesis  in the
type II seesaw mechanism, which introduces vector-like Higgs
triplets, can be the  common source for the matter--anti-matter
asymmetry and the dark matter relic density in Dirac gaugino
models. In order not to deplete the asymmetric dark matter
density,  the Dirac nature of the gauginos has to be maintained
suppressing  Majorana mass terms appropriately, which can be
ensured by imposing an $U(1)_R$ symmetry.   This mechanism can
explain the intriguing coincidence of $\Omega_{DM}/\Omega_B
\approx 5$ with the dark matter mass at the weak scale which can
be naturally obtained by a mild hierarchy of the Higgs triplet
Yukawa couplings. Whereas, in the type I or III leptogenesis induced
by singlet or triplet Majorana fermions, it is not possible to
generate the dark matter asymmetry
as the dark matter number is respected by the decay processes although
they are CP--violating and out--of--equilibrium.

\bigskip

{\bf Acknowledgments:} This work was supported by Korea Neutrino
Research Center through National Research Foundation of Korea
Grant (2009-0083526).


\begin{thebibliography}{99}
\bibitem{note1}
 Note, however, that
the leptogenesis in the type I seesaw introducing right-handed
neutrinos cannot generate a non-vanishing Dirac gaugino dark
matter asymmetry as the DM number is also conserved in the
right-handed neutrino/sneutrino decay.
\end{thebibliography}
\end{document}